\documentclass[aps,twocolumn,floatfix,superscriptaddress]{revtex4-1}
\usepackage[colorlinks=true,linkcolor=blue,anchorcolor=blue,citecolor=blue]{hyperref}
\usepackage{graphicx,bm}

\begin{document}

\title{Site selective 5$f$ electronic correlations in $\beta$-uranium}

\author{Ruizhi Qiu}
\email{qiuruizhi@caep.cn}
\affiliation{Institute of Materials, China Academy of Engineering Physics, Mianyang 621907, Sichuan, China}

\author{Liuhua Xie}
\affiliation{Institute of Materials, China Academy of Engineering Physics, Mianyang 621907, Sichuan, China}
\affiliation{School of Physical Science and Technology, Southwest University, Chongqing 400715, China}

\author{Li Huang}
\email{lihuang.dmft@gmail.com}
\affiliation{Science and Technology on Surface Physics and Chemistry Laboratory, Mianyang 621908, Sichuan, China}

\begin{abstract}
We studied the electronic structure of $\beta$-uranium, which has five non-equivalent atomic sites in its unit cell, by means of the density functional theory plus Hubbard-$U$ correction. We found that the 5$f$ electronic correlations in $\beta$-uranium are moderate. More interesting, their strengths are site selective, depending on the averaged bond length between the present uranium atom and its neighbors. As a consequence, the 5$f$ density of states, occupation matrices, and charge distributions of $\beta$-uranium manifest strongly site dependence.
\end{abstract}

\date{\today}

\maketitle

\section{introduction\label{sec:intro}}

Long time ago, peoples have noticed that some allotropes of light actinides exhibit quite complex crystal structures, which usually have a large number of atoms in their low-symmetry unit cells~\cite{Soderlind1995}. Two celebrated examples are the low-temperature $\alpha$ and $\beta$ phases of plutonium (Pu). Both phases crystallize in the monoclinic structures. As for $\alpha$-Pu, it has 16 Pu atoms within the unit cell and these Pu atoms can be classified into eight non-equivalent types~\cite{Zachariasen1963}. The crystal structure of $\beta$-Pu is even more complex. It has 34 Pu atoms within the unit cell and the non-equivalent types are seven. Thus, a few interesting questions have been raised in the past years, concerning with the lattice structures and electronic structures of Pu and the other light actinides.

First of all, why do the low-temperature phases of light actinides incline to exhibit low-symmetry crystal structures? Now it is well-accepted that the behaviors of the 5$f$ valence electrons of light actinides are key factors to understand this unusual phenomenon~\cite{Moore2009}. In the light actinides (from Th to Pu), the 5$f$ electrons tend to be delocalized and take part in chemical bonding actively~\cite{Huang2020}, in sharp contrast to the late actinides (beyond Pu), in which the 5$f$ electrons are highly localized and chemically inertial at ambient conditions~\cite{Heathman2013,Heathman2005,Roof1980,Huang2019,Huang2020a}. The itinerant 5$f$ electrons usually form narrow and flat bands, leading to very high density of states near the Fermi level. Actually, the specific properties of these flat 5$f$ bands dominate the bonding properties of the light actinides. For example, lattice distortions can split these flat bands and thereby lower the total energy of systems. It is indeed the reason why low-symmetry lattice structures are favored in the low-temperature phases (ground states) of the light actinides~\cite{Soderlind1995,Moore2009}.

Second, since there are multiple non-equivalent actinide atoms in the unit cells, do the site-resolved $5f$ electronic structures differ from each other? More specifically, are the correlation strengths of 5$f$ electrons in these non-equivalent sites close or not? Is that possible to realize the so-called site-selective localized or itinerant 5$f$ states? They are still opening questions. A few years ago, Zhu \emph{et al.} investigated the 5$f$ electronic structures, including site-resolved density of states and hybridization functions, of $\alpha$-Pu by means of a combination of the density functional theory (DFT) and the dynamical mean-field theory (DMFT)~\cite{Zhu2013}. They argued that $\alpha$-Pu's 5$f$ electrons show apparent site dependence. Very recently, Brito \emph{et al.}~\cite{Brito2019} employed the same method (DFT + DMFT)~\cite{Kotliar2006} to study the electronic structures of $\beta$-Pu. They also confirmed the site-selective electronic correlations in $\beta$-Pu, though its site selectivity is less prominent than that in $\alpha$-Pu. We note that the two pioneering works are quite important, but their conclusions are not solid. The authors in their calculations adopted the same interaction parameters (Coulomb repulsion interaction $U$ and Hund's exchange interaction $J_{\text{H}}$) to describe the correlation effects on non-equivalent Pu atoms, which will introduce some sorts of non-consistency. It is not clear whether their conclusions can be generalized to the other 5$f$ electronic systems that have multiple non-equivalent actinide atoms. In other words, we are not sure whether the site selectivity of 5$f$ electronic correlations is a generic feature in the low-symmetry complex actinide metals.

Now let us concentrate on $\beta$-uranium. It might be the most complex allotropes of uranium (U). Since it has more than 30 U atoms in the unit cell under ambient conditions~\cite{Lawson1988}, it is considered as a good testing-bed to examine the site selectivity of 5$f$ electronic correlations. On the other hand, the electronic structures of $\beta$-U were rarely reported~\cite{Li2012,Zhang2017,Beeler2013,Qiu2016,Fast1998}. To the best of our knowledge, the 5$f$ electronic correlations in $\beta$-U has not been well studied. The degree of freedom of 5$f$ localization in this phase remains unknown. In the present work, we intend to investigate the site-resolved 5$f$ electronic structures of $\beta$-U by a first-principles approach. The correlation strengths on non-equivalent U atoms are exactly treated. Our calculated results suggest that similar to $\alpha$-Pu and $\beta$-Pu, $\beta$-U is another example that exhibits site selective 5$f$ electronic correlations.

\begin{figure*}[ht]
\centering
\includegraphics[width=0.95\textwidth]{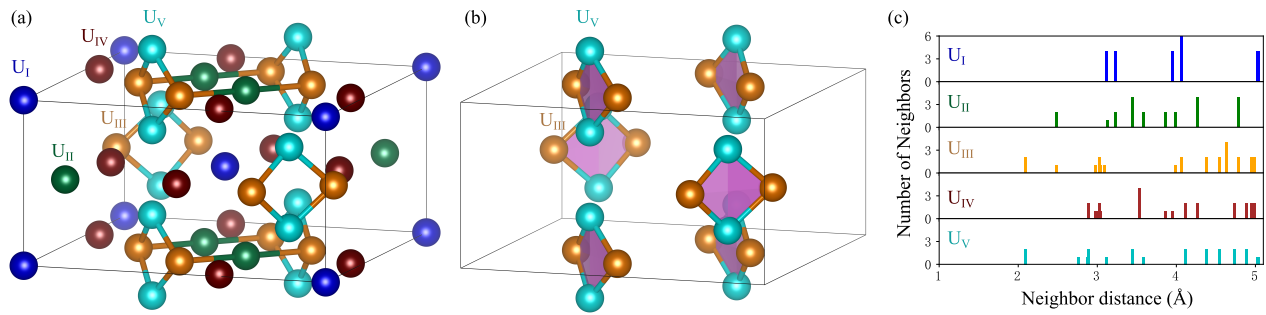}
\caption{(Color online). (a) Schematic crystal structure of $\beta$-U. The uranium atoms at non-equivalent positions (U$_{\rm I}$, U$_{\rm II}$, U$_{\rm III}$, U$_{\rm IV}$, and U$_{\rm V}$) are represented by blue, green, orange, brown, and cyan balls, respectively. The U-U bonds those with bond length $d_{\rm U-U} < 2.5$~\AA~are visualized. (b) Schematic tetrahedra formed by U$_{\rm III}$ and U$_{\rm V}$ atoms. (c) Distribution of interatomic distances for the five crystallographically non-equivalent U atoms in $\beta$-U. The number of neighbors is represented by the height of the lines. \label{fig:struct}}
\end{figure*}

\section{structure\label{sec:struct}}

\begin{table}[b!]
\caption{Lattice parameters of $\beta$-U~\cite{Lawson1988}. Here, $U_{\text{LR}}$ denotes the Coulomb interaction parameter obtained by the linear response approach. See main text for more details. \label{tab:param}}
\begin{ruledtabular}
\begin{tabular}{cclllc}
sites & Wyckoff positions & $x$ & $y$ & $z$ & $U_{\text{LR}}$ \\
\hline
U$_{\rm I}$   & 2$a$  & 0.0    & 0.0    & 0.0    & 2.9~eV \\
U$_{\rm II}$  & 4$f$  & 0.397  & 0.397  & 0.0    & 2.3~eV \\
U$_{\rm III}$ & 8$i$  & 0.5622 & 0.2343 & 0.0    & 1.3~eV \\
U$_{\rm IV}$  & 8$i$  & 0.3655 & 0.0391 & 0.0    & 2.5~eV \\
U$_{\rm V}$   & 8$j$  & 0.1812 & 0.1812 & 0.2556 & 1.4~eV \\
\end{tabular}
\end{ruledtabular}
\end{table}

As the last of the naturally occurring elements in the Periodic Table, U has been extensively studied in the experiment~\cite{Lander1994}. It has three allotropes ($\alpha$, $\beta$, and $\gamma$ phases). The crystal structure of $\beta$-U is the most complex.

As is shown in Fig.~\ref{fig:struct} (a), the lattice of $\beta$-U is tetragonal with space group $P4_2/mnm$ (No.~136)~\cite{Lawson1988}. According to symmetry, the 30 U atoms in the unit cell can be grouped into five non-equivalent types (U$_{\rm I}$, U$_{\rm II}$, U$_{\rm III}$, U$_{\rm IV}$, and U$_{\rm V}$). Their atomic coordinates are collected in Table~\ref{tab:param}. Let us pay close attention to the atomic positions of these non-equivalent U atoms. The vertex and body center of the tetragonal cell are occupied by the U$_{\rm I}$ atoms. The U$_{\rm IV}$ atoms approximately locate at the edge and the face center of the cell. The U$_{\rm II}$, U$_{\rm III}$, and U$_{\rm V}$ atoms form a ``two-bar dumbbell'', in which the center of the bar is occupied by the U$_{\rm II}$ atoms and the distorted square (i.e., the tetrahedron in Fig.~\ref{fig:struct} (b)) is formed by the U$_{\rm III}$ and U$_{\rm V}$ atoms. The existence of these squares implies that the local atomic environment of U$_{\rm III}$ and U$_{\rm V}$ is totally different from that of other U atoms. One can intuitively divide the five non-equivalent U sites into three categories: isolated U$_{\rm I}$ and U$_{\rm IV}$, clustered U$_{\rm III}$ and U$_{\rm V}$, and intermediate U$_{\rm II}$.

We then analyze the local atomic environments for the five non-equivalent U atoms. The distribution of interatomic distances for U$_{\rm I-V}$ is computed and shown in Fig.~\ref{fig:struct} (c). Apparently, the first nearest-neighbor distance ($d_{\text{NN}}$) of U atoms has a wide range. As for U$_{\rm I}$, $d_{\text{NN}}$ is around 3.11~\AA, which is much larger than that of U$_{\rm V}$ and U$_{\rm III}$ ($d_{\text{NN}} \sim $ 2.09~\AA)~\cite{Lawson1988}. As a comparison, $d_{\text{NN}}$ for various non-equivalent Pu atoms in the $\alpha$-Pu lattice only ranges from 2.56 to 2.75~\AA~\cite{Zachariasen1963,Zhu2013}. Thus, it is naturally expected that $\beta$-U could exhibit stronger site selectivity in 5$f$ electronic correlations. In addition, $d_{\text{NN}}$ is related with the strength of electronic correlation. One can expect that the $5f$ electrons of U$_{\rm I}$ and U$_{\rm IV}$ atoms are the most localized. On the contrary, the 5$f$ electrons of U$_{\rm III}$ and U$_{\rm V}$ atoms are the most itinerant. The itinerancy or localization of 5$f$ electrons of U$_{\rm II}$ atoms are intermediate.

\section{methods\label{sec:method}}

Note that the DFT + DMFT method might be the most powerful approach ever established to study the electronic structures and physical properties of actinides~\cite{Kotliar2006}. It has been widely employed to study Pu and its compounds~\cite{Zhu2013,Huang2019,Huang2020a}. However, it requires huge computer resources and is extremely time-consuming. If there are multiple non-equivalent atoms in the unit cell, the situation becomes even worse. The reason is that each non-equivalent atom is described by a multi-orbital quantum impurity model, which should be solved by quantum impurity solvers repeatedly in the DMFT iterations. In the previous works about $\beta$-Pu, a simplified quantum impurity solver based on the one-crossing approximation was used~\cite{Brito2019}, with which the computational accuracy is restricted. In the present work, we adopted a traditional but more economic approach, namely DFT + $U$, to study the electronic structure of $\beta$-U~\cite{Liechtenstein1995,Dudarev1998}. This approach has been successfully applied to a large amount of actinide compounds~\cite{Savrasov2000,Shick2005,Shick2006,Jomard2008,Dorado2009}.

All the electronic structure calculations are performed using Vienna {\it Ab initio} Simulation Package (VASP)~\cite{Kresse1996}. We consider the experimental lattice structure of $\beta$-U only~\cite{Lawson1988}. The relativistic effect, i.e spin-orbit coupling, is included in all the calculations. However, the magnetic moment for each U atom is assumed to be zero. As for the exchange-correlation term in the DFT Hamiltonian, the Perdew-Burke-Ernzerhof functional (generalized gradient approximation) is used~\cite{Perdew1996}. The Kohn-Sham equation is solved within the projector augmented wave formalism~\cite{Blochl1994,Kresse1999}. The valence electronic configuration for U atom is $6s^26p^66d^15f^37s^2$. From the convergence tests, the optimal cutoff energy for plane wave basis is 450 eV and the division of Monkhorst-Pack $k$-mesh is $3\times3\times7$.

The 5$f$ electronic correlations in U atoms are important. We used a static mean-field scheme (i.e. the Hubbard-$U$ correction method) to capture their contributions. According to the formalism of DFT + $U$, an on-site interaction term is added to the Hamiltonian, which reads
~\cite{Dudarev1998}:
\begin{eqnarray}
E_{U}[\{{\bm n}^J\}] = \sum_{J}\frac{U^J}{2}{\rm Tr}\left[{\bm n}^J(1-{\bm n}^J)\right].
\label{eq:dudarev}
\end{eqnarray}
Here $J$ is the index of correlated atom, $U^J$ is the effective Coulomb interaction, and ${\bm n}^J$ is the occupation number matrix. ${\bm n}^J$ can be evaluated by the projection of Kohn-Sham orbitals ($\psi_{{\bm k},\nu}$) into the states of $5f$ localized orbitals ($\phi^J_m$):
\begin{eqnarray}
n^J_{mm'} = \sum_{{\bm k},\nu}f_{{\bm k},\nu}\left\langle \phi^J_{m} | \psi_{{\bm k},\nu} \right\rangle \left\langle  \psi_{{\bm k},\nu} | \phi^J_{m'} \right\rangle,
\end{eqnarray}
 where ${\bm k}$, $\nu$, and $m$ are Bloch wavevector, band index, and index of localized orbitals, respectively, and $f_{{\bm k},\nu}$ is the Fermi-Dirac distribution of the Kohn-Sham states. For $\beta$-U, the values of $U^J$ for the five non-equivalent atoms should be determined using the linear response approach (i.e. $U^{J}_{\rm LR}$)~\cite{Cococcioni2005,Qiu2020}. Then in the successive calculations, we considered four different cases: $U^J = U^{J}_{\rm LR}$, 0.0~eV, 2.0~eV, and 3.0~eV.

The occupation number matrix ${\bm n}^{J}$ is Hermite and the eigenvalues of this matrix are the occupation number $\{{\bar n}^{J}_m\}$, which could be used to determine the degree of $5f$ electronic localization or delocalization. In terms of $\{{\bar n}^{J}_m\}$, the Hubbard-$U$ functional in equation (\ref{eq:dudarev}) could be transformed as $\sum_{J,m}{U^J}{\bar n}^{J}_m(1-{\bar n}^{J}_m)/2$. Clearly, the fully localized electrons (${\bar n}^{J}_m \sim 1$) and empty occupation (${\bar n}^{J}_m\sim 0$) are energetically favorable. Thus, the Hubbard-$U$ correction scheme favors Mott localization (i.e. insulating feature), instead of fractional occupation of localized orbitals (i.e. the metallic-like hybridization). We can use the following quantity to measure the degree of $5f$ delocalization for each correlated atom:
\begin{eqnarray}
{\cal D}^J = \sum_{m}{\bar n}^{J}_m(1-{\bar n}^{J}_m).\label{eq:degree}
\end{eqnarray}
If the $5f$ electrons tend to be fully localized, ${\bar n}^{J}_m \sim 1$ and ${\cal D}^J \sim 0$. On the contrary, if the $5f$ electrons favor itinerant states, ${\bar n}^{J}_m \sim 1$ is far away from 1.0 and ${\cal D}^J$ is much greater than $0$.

During the practical DFT+$U$ calculations, electronic metastable states often emerge. For the system with various interaction parameters, the problem is more worse. First, we employ the occupation matrix control scheme~\cite{Jomard2008,Dorado2009} with only a single occupation matrix enumerated and $U$-ramping scheme~\cite{Meredig2010} with increment less than 0.1 eV. Then we apply a new scheme by adding a local perturbation $\sum_{m}\alpha^J_m|\phi^J_m\rangle\langle\phi^J_m|$ to the external potential, and perform a first calculation with a random set of $\{\alpha^J_m\}$ and a second calculation without local perturbation. A dozen sets of random number are tried and the lowest energy state of this scheme repeatedly appear, which confirm the validity of this method. The resulting lowest energy from these calculations is much lower than those from occupation matrix control and $U$-ramping scheme. It is suggested that the lowest-energy DFT+$U$ state of $\beta$-U could be approached by employing these schemes.

\section{results\label{sec:result}}

\begin{figure}[ht]
\centering
\includegraphics[width=\columnwidth]{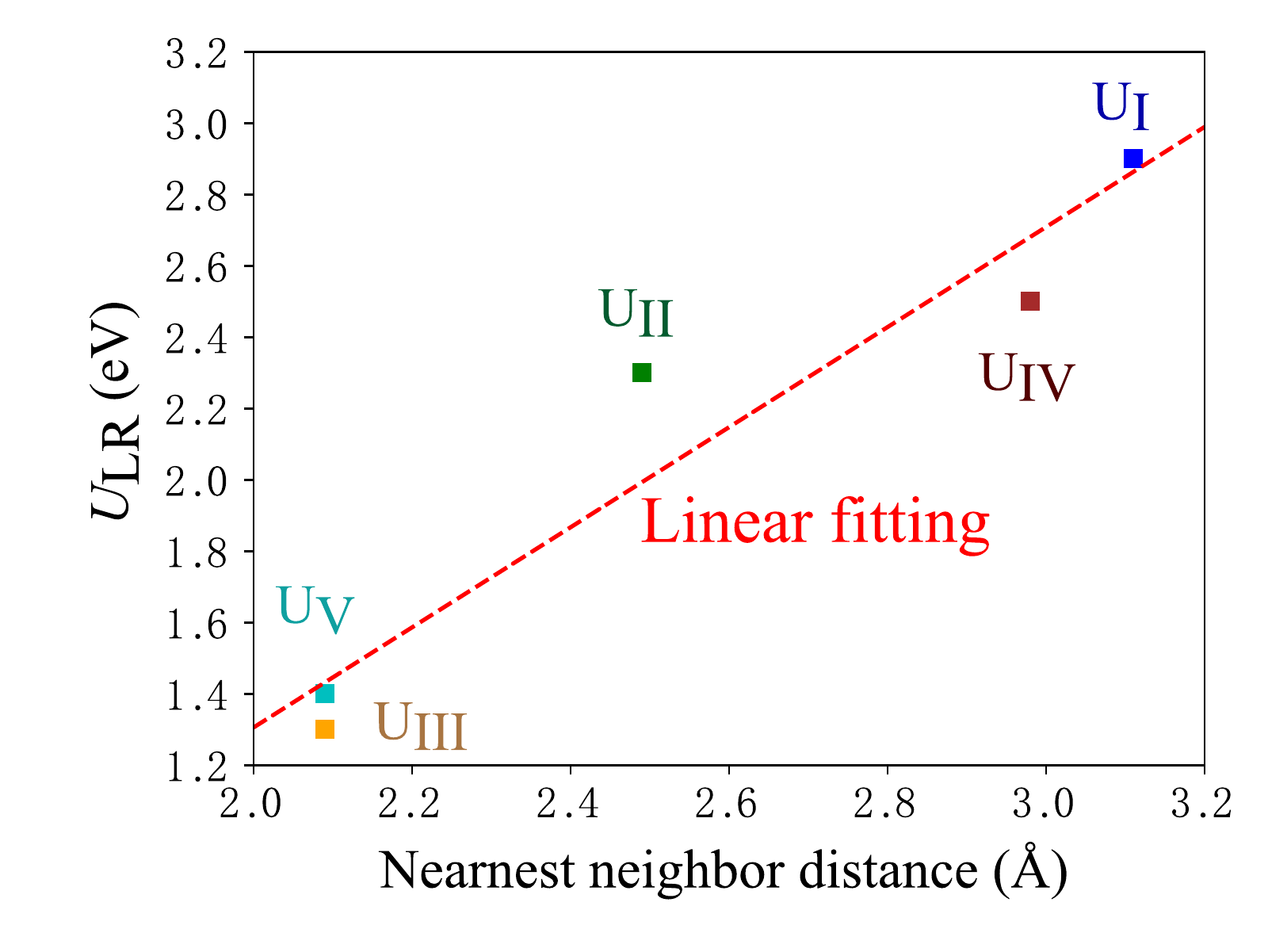}
\caption{The interaction parameters $U_{\rm LR}$ from linear response calculation versus the nearest neighbor distance $d_{\rm NN}$. The dashed line implies the linear dependence of $U_{\rm LR}$ with respect to $d_{\rm NN}$.\label{fig:ulr}}
\end{figure}

\begin{figure*}[ht]
\centering
\includegraphics[width=0.99\textwidth]{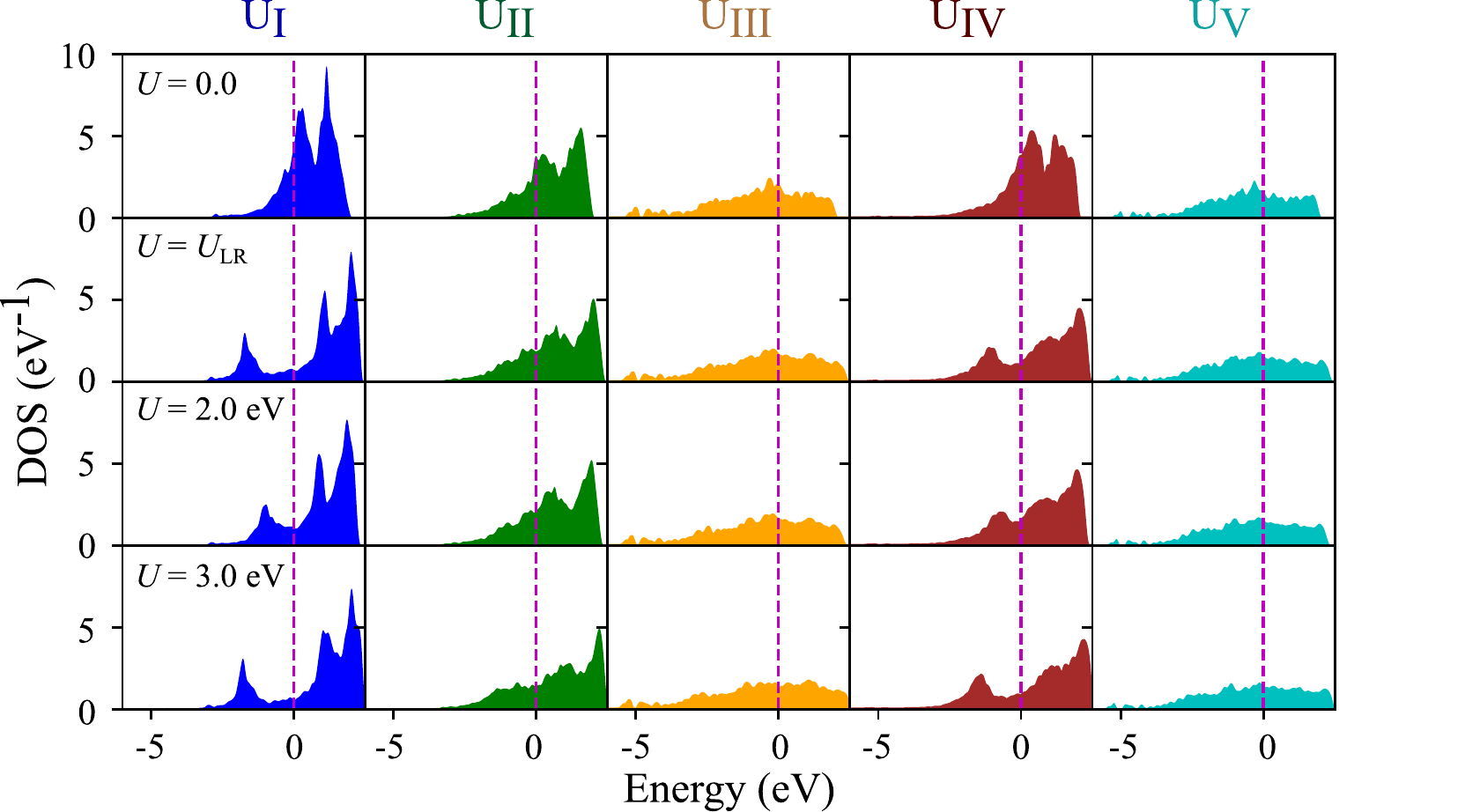}
\caption{The 5$f$ partial density of states for the five non-equivalent U atoms (from left to right) of $\beta$-U obtained by DFT + $U$ calculations. Four different cases are considered. They are $U^J$ = 0.0, $U_{\rm LR}$, 2.0, and 3.0 eV (from top to bottom). The vertical dashed line denotes the Fermi level. \label{fig:dos}}
\end{figure*}

\begin{figure*}[ht]
\centering
\includegraphics[width=0.95\textwidth]{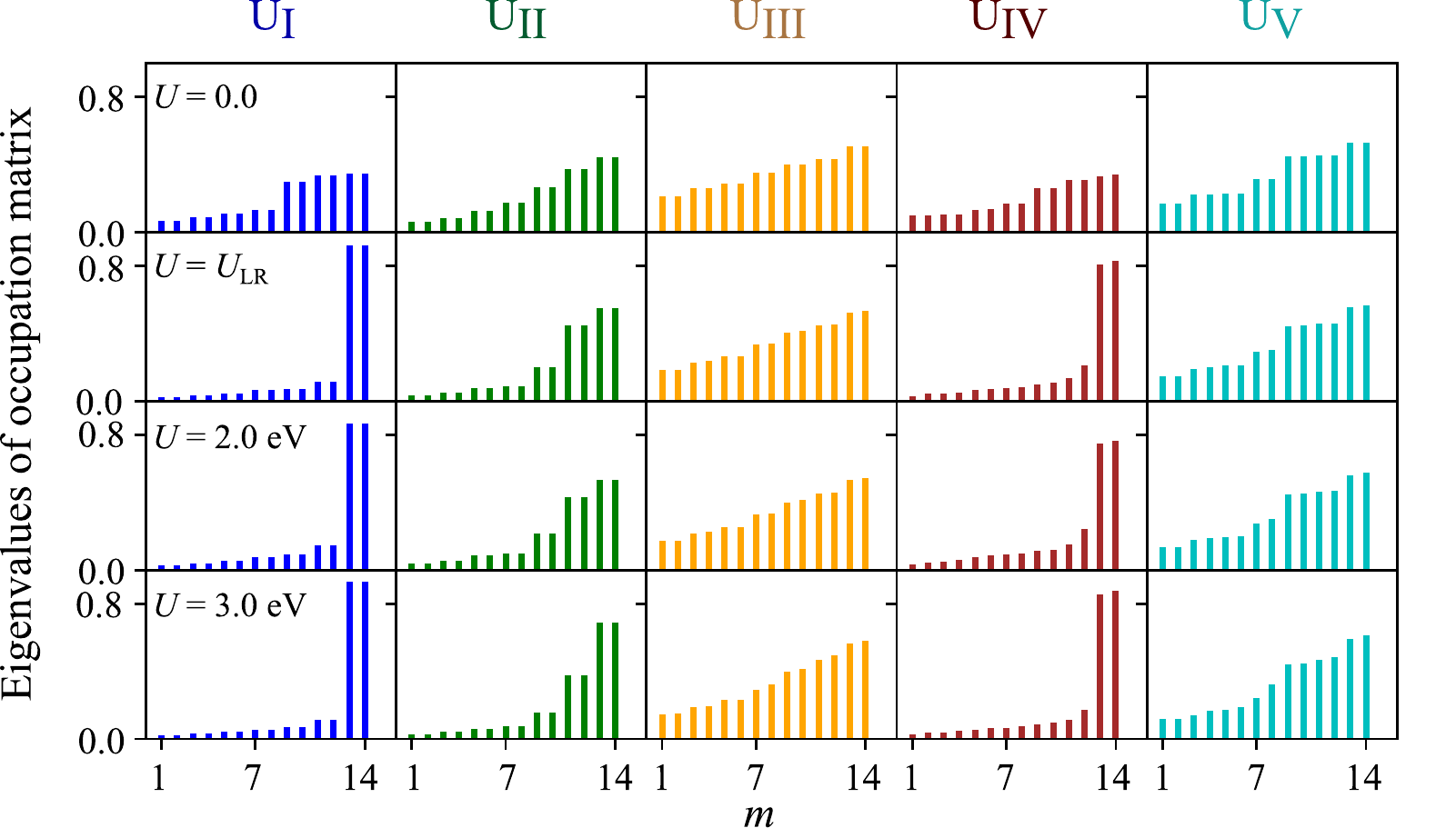}
\caption{The eigenvalues of occupation number matrix ${\bar n}^I_{m}$ for the five non-equivalent U atoms (from left to right) of $\beta$-U obtained by DFT + $U$ calculations. Four different cases are considered. They are $U^J$ = 0.0, $U_{\rm LR}$, 2.0, and 3.0 eV (from top to bottom). \label{fig:eigen}}
\end{figure*}

\begin{figure}[ht]
\centering
\includegraphics[width=\columnwidth]{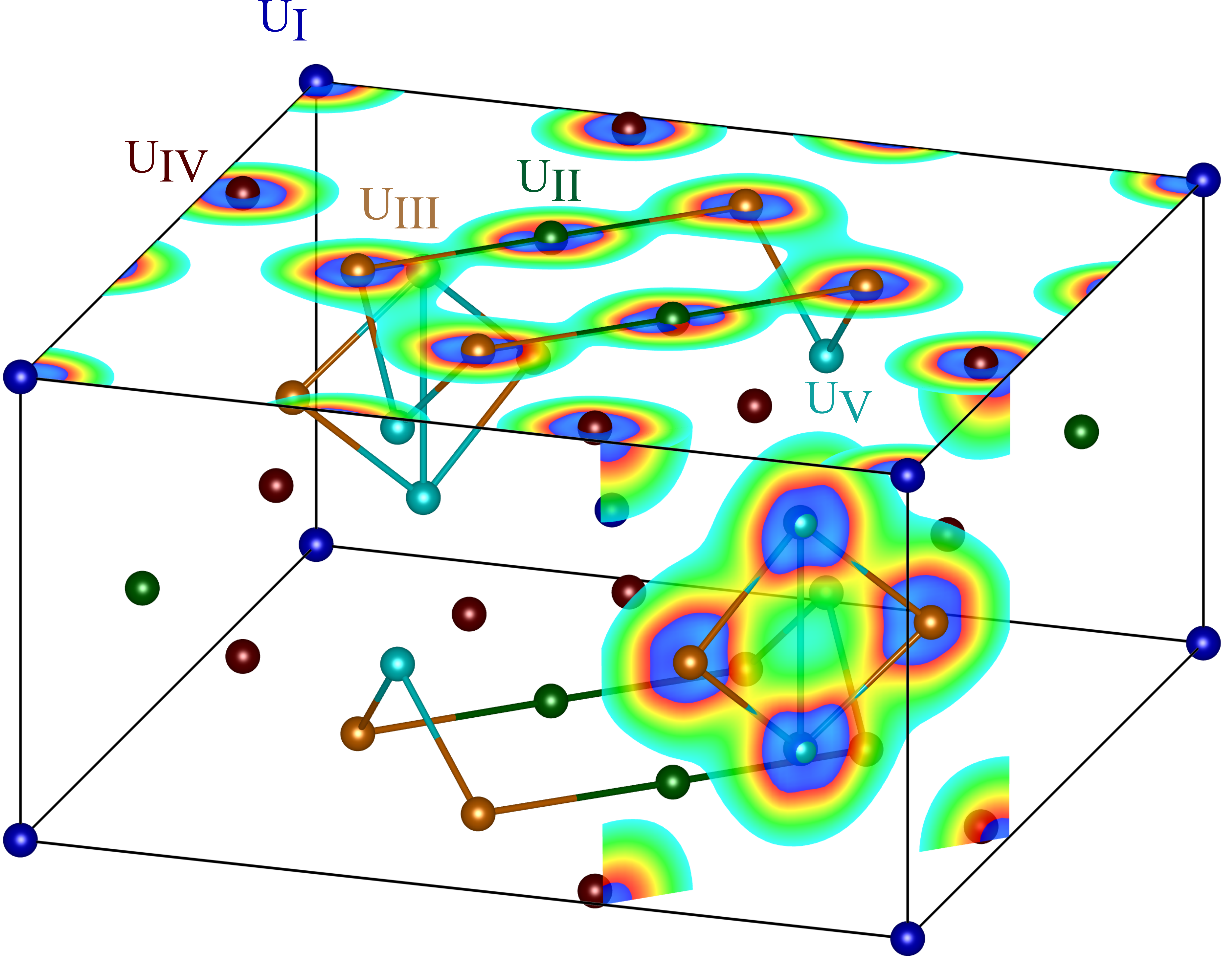}
\caption{Charge distributions of $\beta$-U obtained by the DFT + $U_{\text{LR}}$ calculation. In this figure, only the charge densities in two planes are visualized. The (001) plane contains the U$_{\rm I}$ $\sim$ U$_{\rm IV}$ atoms, while another plane contains the U$_{\rm III}$ $\sim$ U$_{\rm V}$ atoms. \label{fig:charge}}
\end{figure}

\emph{Site-resolved interaction parameters.} At first, we employed the linear response approach to determine the interaction parameters $U^{J}$ for the five non-equivalent U atoms~\cite{Cococcioni2005,Qiu2020}. The calculated results are collected in Table~\ref{tab:param}. We note that these results are obtained by using the PBE exchange-correlation functional. We also examined the other exchange-correlation functionals, such as the CA-PZ functional which is based on the local density approximation~\cite{Ceperley1980,Perdew1981} and variation of the PBE functional (the PBEsol functional)~\cite{Perdew2008}. The obtained results are nearly unchanged. These benchmark calculations suggest that the choice of exchange-correlation functional does not affect the final results. Our calculations are reliable and the dominant effect is indeed the local atomic environment.

We find that the U$_{\rm I}$ atoms have the largest $U_{\text{LR}}$ (= 2.9~eV), which means that their 5$f$ electrons are more localized than the others. On the contrary, the U$_{\rm III}$ atoms have the smallest $U_{\text{LR}}$ (= 1.3~eV), which means that their 5$f$ electrons are more itinerant and more easily hybridized with conduction electrons than the others. These results also suggest that the interaction parameter has the same sequence with the nearest neighbor distance $d_{\text{NN}}$. The further away from the nearest neighbor, the larger the interaction parameter is. The tendency is shown in Fig.~\ref{fig:ulr} in which $U_{\rm LR}$ as a function of $d_{\text{NN}}$ is plotted. In addition, the linear dependence of $U_{\rm LR}$ with respect to $d_{\rm NN}$ is found from the fitting curve in Fig.~\ref{fig:ulr}. Note that the categories as proposed in Section~\ref{sec:struct} are confirmed.

\emph{$5f$ partial density of states.} In Fig.~\ref{fig:dos}, the calculated $5f$ partial density of states for the five non-equivalent U atoms are shown. Both DFT and DFT+$U$ ($U^J$ = $U_{\rm LR}$, 2.0, and 3.0 eV) are considered. Obviously, the spectra for U$_{\rm III}$ and U$_{\rm V}$ atoms are quite different from those of the other atoms, irrespective of the interaction strength $U^{J}$ or the used exchange-correlation functionals. Compared to the spectra of U$_{\rm I}$, U$_{\rm II}$, and U$_{\rm IV}$ atoms, those of U$_{\rm III}$ and U$_{\rm V}$ atoms resemble big ``humps''. They exhibit more small peaks, lower intensities at the Fermi level, and larger effective bandwidth. The appearance of the multi-peak feature indicates that there is a strong hybridization between $5f$ and $spd$ conduction electrons. The spectra for U$_{\rm I}$ and U$_{\rm IV}$ atoms look quite similar. There are lower and upper Hubbard bands, which are shown as the two broad peaks below and above the Fermi level. In particular, the $5f$ spectra of U$_{\rm I}$ atoms have a new peak at the lower Hubbard band. This is a remarkable feature of the $5f$ electron localization. Similar feature is found for the U$_{\rm IV}$ atoms but the peak at the lower Hubbard band is lower. For U$_{\rm II}$ atoms, their spectra are more like those of U$_{\rm I}$ and U$_{\rm IV}$ atoms. But the Hubbard bands are shifted toward the Fermi level slightly. In Fig.~\ref{fig:dos}, we also examine the influence of interaction strength. For U$_{\rm III}$ and U$_{\rm V}$ atoms, their spectra change a little when $U^{J}$ varies. However, for U$_{\rm I}$, U$_{\rm II}$, and U$_{\rm IV}$ atoms, their spectra are easily affected by $U^{J}$. When $U^{J}$ is increased, the Hubbard bands are pushed away from the Fermi level, and the spectral weights near the Fermi level are reduced.

Comparing with the partial density of states of $\alpha$-Pu~\cite{Zhu2013} and $\beta$-Pu~\cite{Brito2019}, we find that the spectra of $\beta$-U exhibit more site-resolved features. Clearly, from the $5f$ partial density of states, it is no doubt that there ares apparent site selective electronic correlations in $\beta$-U and the site selectivity in $\beta$-U might be stronger than those in $\alpha$-Pu and $\beta$-Pu.

\emph{5$f$ orbital occupations.} Figure~\ref{fig:eigen} shows the eigenvalues of occupation matrix from DFT + $U$ calculation. For the 14 eigenvalues ${\bar n}^{I}_m$ of U$_{\rm I}$ atoms, two of these approximately equal to $1$ ($m = 13$ and 14) and the others are close to $0$. This indicates that the $5f$ electrons of U$_{\rm I}$ atoms are extremely localized. The ${\bar n}^{I}_m$ of U$_{\rm IV}$ atoms has a similar feature but the deviation from $1/0$ is a bit larger. Relatively, the $5f$ electrons of U$_{\rm IV}$ atoms are also localized but the degree of localization is somewhat lower. For the other atoms, the $5f$ electron delocalization is outstanding since most of ${\bar n}^I_m$ are close to $1/2$. Using the definition Eq.~(\ref{eq:degree}), the degree of delocalization ${\cal D}^I$ from DFT + $U_{\rm LR}$ is $0.8$, $1.8$, $3.0$, $1.2$, $2.8$ for U$_{\rm I}$ $\sim$ U$_{\rm V}$ atoms, respectively. Our intuitive expectation is confirmed again here. The $5f$ electrons around the U$_{\rm I}$ atoms are the most localized and those around the U$_{\rm III}$ atoms are the most delocalized. The degree of localization and the effective interaction have the same sequence: U$_{\rm I}$ $>$ U$_{\rm IV}$ $>$ U$_{\rm II}$ $>$ U$_{\rm V}$ $>$ U$_{\rm III}$.

\emph{Distribution of charge density.} The site-selective electronic localization among the five non-equivalent U atoms could also be identified in the charge density of $\beta$-U. In Fig.~\ref{fig:charge}, the charge distribution in (001) plane and the optimal plane for the distorted square is plotted. The charge distributions among the U$_{\rm II}$, U$_{\rm III}$, and U$_{\rm V}$ atoms are very clear while the charge distributions around U$_{\rm I}$ and U$_{\rm IV}$ atoms are isolated. It is suggested that the 5$f$ electrons in U$_{\rm I}$ and U$_{\rm IV}$ atoms favor localized states. While those in the U$_{\rm II}$, U$_{\rm III}$, and U$_{\rm V}$ atoms incline to be itinerant, thus strong chemical bonding behaviors among these U atoms are observed.

\section{conclusion\label{sec:summary}}

In this work, we study the 5$f$ electronic structures of $\beta$-uranium ($\beta$-U) by using the DFT and DFT + $U$ approaches. We determine the site-resolved interaction parameters, and confirm the existence of site-selective electronic correlation in $\beta$-U. For the atoms in the vertex and center of the cell (U$_{\rm I}$ and U$_{\rm IV}$ atoms), the $5f$ electrons is strongly correlated. While for the atoms in the distorted square (U$_{\rm II}$, U$_{\rm III}$, and U$_{\rm V}$ atoms), the $5f$ electrons are highly itinerant. The degree of localization on each position is closely related to the local atomic environment. This conclusion is supported by the analysis of the neighbor distance, the $5f$ partial density of states, and the charge distribution of $\beta$-U. Obviously, this site-selective electronic correlation might be a generic feature for those actinide metals which have multiple non-equivalent sites in the unit cell. Thus, we believe that the high-pressure phases of Am~\cite{Roof1980}, Cm~\cite{Heathman2005,Huang2020a}, and Cf~\cite{Heathman2013,Huang2019} probably exhibit some sorts of site selectivity. Further theoretical works are highly desired.

\begin{acknowledgments}
This work is funded by National Natural Science Foundation of China (Grant No.~11874329, 11934020, and 22025602).
\end{acknowledgments}

\bibliography{beta}
\end{document}